\documentclass[11pt,english]{article}
\usepackage[T1]{fontenc}
\usepackage[latin9]{inputenc}
\usepackage{amsmath}

\makeatletter
\usepackage{amsmath}
\usepackage{amssymb}
\usepackage{latexsym}
\usepackage{epsfig}
\usepackage[numbers,sort&compress]{natbib}

\newcommand{\be}{\begin{equation}}
\newcommand{\ee}{\end{equation}}

\allowdisplaybreaks

\makeatother

\usepackage{babel}
\begin{document}


\vskip 3.0cm

\centerline{\Large \bf Superluminal neutrinos and domain walls}
\vspace*{6.0ex}
\centerline{\large Peng Wang, Houwen Wu and Haitang Yang}
\vspace*{4.0ex}
\centerline{\large \it  School of Physical Electronics}
\centerline{\large \it University of Electronic Science and Technology of China}
\centerline{\large \it Chengdu, 610054, China} \vspace*{1.0ex}
\centerline{pengw@uestc.edu.cn, iverwu@uestc.edu.cn, hyanga@uestc.edu.cn}
\bigskip
\smallskip

\begin{abstract}
In this letter, we propose that the recent measurement of superluminal neutrinos
in OPERA could be explained by the existence of a domain wall which is left
behind after the phase transition of some scalar field in the universe. The
scalar field couples to the neutrino and photon field with different effective
couplings. It causes different effective metrics and the emergence of
superluminal neutrinos.  Moreover, if the
supernova and the earth are in the same plane parallel to the wall, or the
thickness of the wall is much smaller than the distance from the supernova to
the earth, the contradiction between OPERA and SN1987a can be reconciled.
\end{abstract}
\vspace*{10.0ex}

Recently OPERA collaboration \cite{:2011zb} published their results which
show that the muon neutrino moves faster than the speed of light ($c$).
The $\nu_{\mu}$ arrives at the Gran Sasso laboratory from CERN by $60$ ns
earlier than the photon, with a distance around $730$ km. The beam of
$\nu_\mu$ has mean energy $17$ GeV. The measured relative amount of
the superluminal velocity is:
\begin{equation}
\frac{v_\nu-c}{c}=(2.48\pm 0.28({\rm stat}) \pm 0.30({\rm sys})) \times 10^{-5},
\end{equation}
with a statistical significance of $6.0\,\sigma$. Many theoretical
explanations have been proposed immediately, respecting or
violating Lorentz invariance \cite{superluminal2011}.
While as early as in 2005, there were discussions on the
possibility of superluminal neutrinos \cite{earlier superluminal
neutrinos}.

As one of the cornerstones of special relativity, the constancy of speed
of light is from the lessons in electromagnetic theory. It is reasonable
that all the participants of $U(1)$ gauge interaction respect the speed
limit $c$. However, since neutrinos only play roles in weak and gravitional
interactions, there exit possibilities of superluminal propagation.
This property serves as an ingredient of the $SU(2)$ symmetry breaking.

In this letter, we propose a domain wall model explicitly breaking
the $SU(2)$ symmetry while spontaneously breaking Lorentz invariance.
In the literature, there are discussions on variation of the light speed
caused by domain walls (brane-worlds) \cite{speed varying}.
In a companion paper, we address the possible influences on superluminal neutrinos
from two other topological defects, cosmic strings and monopoles \cite{Wang:2011oc}.
To simplify the story, only a real
scalar field $\phi$, responsible to generate a domain wall where we
live, a Dirac neutrino field $\psi$ and the photon field $A_{\mu}$ are
included in the model. We also assume the neutrino is massless since its mass
is very tiny compared to the its energy.
Since the gravitational field is very weak around the earth  it is safe to
consider our model in Minkowski space.
The effective Lagrangian is given by%
\begin{align}
\mathcal{L}  & \mathcal{=}i\overline{\psi}\gamma^{\mu}\partial_{\mu}\psi
-\frac{1}{4}F_{\mu\nu}F^{\mu\nu}+\frac{1}{2}\left(  \partial_{\mu}\phi\right)
^{2}-\frac{1}{4}\lambda\left(  \phi^{2}-\sigma^{2}\right)  ^{2}%
\label{effective lagrangian}\\
& +\frac{ig}{M^{4}}\overline{\psi}\gamma_{\mu}\partial_{\nu}\psi\partial^{\mu
}\phi\partial^{\nu}\phi-\frac{g^{\prime}}{8M^{4}}\partial^{\mu}\phi
\partial^{\nu}\phi F_{\mu\rho}F_{\nu}^{\rho}+\cdots,\nonumber
\end{align}
where $M$ is the mass scale where new physics arise. $g$ and $g^{\prime}$ are the
couplings for the effective operators with the order of unity in the
absence of fine tuning in the new physics. $Z_{2}$ symmetry is imposed on the
Lagrangian in the construction of the effective operators. Therefore, the lower dimensional
operators, $\overline{\psi}\gamma_{\mu}D_{\nu}\psi\partial^{\mu}\partial^{\nu
}\phi$ and $\partial^{\mu}\partial^{\nu}\phi F_{\mu\rho}F_{\nu}^{\rho}$ are
excluded in the effective Lagrangian due to the $Z_{2}$ symmetry. The two
eight dimensional operators in the second line of
eqn. (\ref{effective lagrangian}) are the lowest dimensional operators which are
relevant to modifying the kinetic terms of neutrino and photon fields. $\cdots$
are either irrelevant effective operators or higher dimensional
operators.\ $i$ is put in front of $g$ to make $g$ real.

After the big bang, the scalar field $\phi$ is assumed to go through phase
transition from the symmetric phase $\phi=0$ to the broken one $\phi=\pm\sigma.$ The Kibble
mechanism \cite{Kibble Mechanism} tells us that productions of various
topological effects, such as domain walls, cosmic strings and magnetic
monopoles, are unavoidable in the early universe. Suppose there is a domain
wall produced by a scalar field around our earth. The wall is located in the
$xy$ plane at $z=0.$ The profile of the domain wall
\begin{equation}
\phi_{w}\left(  z\right)  =\sigma\tanh\left(  \frac{z}{\Delta}\right),
\label{domain wall profile}%
\end{equation}
is determined by the equation of motion for $\phi$ with boundary conditions
$\phi_{w}\left(  \pm\infty\right)  =\pm\sigma$.
$\Delta=\left(  \frac{\lambda}{2}\right)  ^{-\frac{1}{2}}\sigma^{-1}$
characterizes the thickness of the wall.

Since the wall is around the earth, its existence effectively modifies the
neutrino's kinetic term as
\begin{equation}
i\left(  \eta_{\mu\nu}+\frac{g}{M^{4}}\partial^{\mu}\phi_{w}\partial^{\nu}%
\phi_{w}\right)  \overline{\psi}\gamma_{\mu}\partial_{\nu}\psi.
\end{equation}
Therefore, the equation of motion for $\psi$ is
\begin{equation}
\left(  \eta_{\mu\nu}+\frac{g}{M^{4}}\partial^{\mu}\phi_{w}\partial^{\nu}%
\phi_{w}\right)  \gamma_{\mu}\partial_{\nu}\psi+\cdots=0,
\end{equation}
where $\cdots$ are terms involving $\widetilde{\phi}=\phi-\phi_{w}$ and have
no relevance to our discussion. With the help of eqn. (\ref{domain wall profile}%
), the effective metric the neutrino see is
\begin{equation}
ds^{2}=-dt^{2}+dx^{2}+dy^{2}+\left(  1+\frac{g\sigma^{2}}{M^{4}\Delta^{2}%
\cosh^{4}\left(  \frac{z}{\Delta}\right)  }\right)  dz^{2}.
\end{equation}

It is interesting to note that the effective light cone for the neutrino get
modified in $z$-direction only. It indicates that the massless neutrino travels at the
same speed of light in $x$- and $y$-directions, while in $z$-direction the
neutrino travels faster than light due to the existence of the factor $\frac{g\sigma^{2}%
}{M^{4}\Delta^{2}\cosh^{4}\left(  \frac{z}{\Delta}\right)  }$. In order to put
the strictest lower bound on $\frac{g\sigma^{2}}{M^{4}\Delta^{2}}$, we
consider a neutrino traveling from $\left(  x,y,z_{i}\right)  $ to$\left(
x,y,z_{f}\right)  $ along $z$-direction. In this situation, the domain wall causes the
greatest deviation between the speed of a neutrino and that of light. In this
scenario, the time for the neutrino to travel from $\left(  x,y,z_{i}\right)
$ to $\left(  x,y,z_{f}\right)  $ is simply given by
\begin{align}
t_{\nu}  & =\int_{z_{i}}^{z_{f}}\sqrt{1+\frac{g\sigma^{2}}{M^{4}\Delta
^{2}\cosh^{4}\left(  \frac{z}{\Delta}\right)  }}dz\\
& \approx\int_{z_{i}}^{z_{f}}\left(  1+\frac{g\sigma^{2}}{2M^{4}\Delta
^{2}\cosh^{4}\left(  \frac{z}{\Delta}\right)  }\right)  dz\nonumber\\
& =\left(  z_{f}-z_{i}\right)  +\frac{g\sigma^{2}}{6M^{4}\Delta}\left(
f\left(  \frac{z_{f}}{\Delta}\right)  -f\left(  \frac{z_{i}}{\Delta}\right)
\right),  \nonumber
\end{align}
where $f\left(  z\right)  =\left(  2+\frac{1}{\cosh^{2}\left(  z\right)
}\right)  \tanh\left(  z\right)  $. It is expected, as we will show
later, that $\frac{g\sigma^{2}}{M^{4}\Delta^{2}}\ll1$\ as a consequence of $\delta
=\frac{v_{\nu}-c}{c}\sim10^{-5}$ measured in OPERA. Thus higher order
terms of $\left(  \frac{g\sigma^{2}}{M^{4}\Delta^{2}}\right)  $\ in the
second line of the above equation are discarded.

Parallel to the neutrino calculation, the photon field also undergoes an effective metric
produced by the effective Lagrangian. It is straightforward to write down the
effective metric as
\[
ds^{2}=-dt^{2}+dx^{2}+dy^{2}+\left(  1+\frac{g^{\prime}\sigma^{2}}{M^{4}%
\Delta^{2}\cosh^{4}\left(  \frac{z}{\Delta}\right)  }\right)  dz^{2}.%
\]

\noindent The time for a photon traveling from $\left(  x,y,z_{i}\right)  $ to $\left(
x,y,z_{f}\right)$  is

\[
t_{c}\approx\left(  z_{f}-z_{i}\right)  +\frac{g^{\prime}\sigma^{2}}%
{6M^{4}\Delta}\left(  f\left( \frac{ z_{f}}{\Delta}\right)  -f\left(
\frac{z_{i}}{\Delta}\right)  \right).
\]

\noindent The relative difference of the neutrino speed with respect to the speed of
light is
\begin{equation}
\delta=\frac{v_{\nu}-c}{c}=\frac{t_{c}-t_{\nu}}{t_{\nu}}\approx\frac{\left(
g^{\prime}-g\right)  \sigma^{2}}{6M^{4}\Delta^{2}}\frac{f\left(  \frac{z_{f}%
}{\Delta}\right)  -f\left(  \frac{z_{i}}{\Delta}\right)  }{\frac{z_{f}}%
{\Delta}-\frac{z_{i}}{\Delta}}.\label{delta expression}%
\end{equation}
$f\left(  z\right)  $ is a monotonically increasing function so $\delta$ is always
positive as long as $g<g^{\prime}.$\ The effective couplings $g$ and
$g^{\prime}$ are determined by the new physics beyond $M$. If there is no
symmetry or fine-tuning in the new physics regime, $g,$ $g^{\prime}$ and
$g-g^{\prime}$ should be the order of unity.

It is subtle to calculate physical distances in any experiment since different
fields see different effective metrics in our model. Specifically, the way the
coordinates $z_{f}$ and $z_{i}$ related to physical distances measured in
various experiments, such as OPERA or SN1987a, depends on the kind of fields
employed in the measurements of physical distances. The neutrino baseline
length in OPERA is obtained by analyzing the GPS benchmark positions. The
distance to SN1987a is calculated using the observed angular size of it rings
\cite{SN1987a distance}. Thus in both experiments the photon's effective
metric have been used to calculate physical distances. However, the
differences between physical distances and $\left\vert z_{f}-z_{i}\right\vert
$ are $\mathcal{O}\left(  \frac{\sigma^{2}}{M^{4}\Delta^{2}}\right)  $ if
$g\prime$ is the order of unity. Only keeping the leading order terms in
$\delta$, one can simply treat
$\left\vert z_{f}-z_{i}\right\vert \ $ as physical distances in the following discussion.

Nonzero $\delta$ can be used to determine $\frac{\left(  g^{\prime}-g\right)
\sigma^{2}}{M^{4}\Delta^{2}}$ as long as $z_{i}$ and $z_{f}$ are known. In
order to explain OPERA results, $\frac{f\left(  z_{f}^{\text{OPERA}}%
/\Delta\right)  -f\left(  z_{i}^{\text{OPERA}}/\Delta\right)  }{z_{f}%
^{\text{OPERA}}/\Delta-z_{i}^{\text{OPERA}}/\Delta}$ ought to be the order of
unity. It means that $z_{f}^{\text{OPERA}}$\ and $z_{i}^{\text{OPERA}}%
$\ have to be in the range of $\left(  -\Delta,\Delta\right)  $. The global
maximum of $\left(  f\left(  \frac{z_{f}}{\Delta}\right)  -f\left(
\frac{z_{i}}{\Delta}\right)  \right)  /\left(  \frac{z_{f}}{\Delta}%
-\frac{z_{i}}{\Delta}\right)  $,which is $3$, is achieved as $z_{f}\rightarrow
z_{i}=0$. Hence the measurement of $\delta$ can be used to put a lower bound
on $\frac{\sigma^{2}}{M^{4}\Delta^{2}}$ even without detailed knowledge of
$z_{i}$ and $z_{f}$. The measurements $\delta_{\text{OPERA}}$ in OPERA implies%
\begin{equation}
\frac{\left(  g^{\prime}-g\right)  \sigma^{2}}{M^{4}\Delta^{2}}\sim
\frac{\sigma^{2}}{M^{4}\Delta^{2}}\gtrsim10^{-5}, \label{domain condition}%
\end{equation}
where the fact that $g^{\prime}-g$ is the order of unity in the absence of
fine-tuning has been used.

It is interesting to ask what is the thickness of the domain wall $\Delta$
provided its existence could explain OPERA results. Three possibilities for
$\Delta$ are proposed as follow,

\begin{enumerate}
\item The peculiar velocity of the domain wall is zero. The peculiar velocity
of the sun with respect to distant galaxies\ is estimated to be $\sim 400$%
km/sec \cite{Peculiar Velocity of the Sun}. The data of superluminal neutrinos
have been taken from\ 2009 to 2011 in OPERA. During the period, the sun travelled $3.8\times
10^{10}$ km $\sim10^{-3}$ ly. In order to explain the superluminal neutrinos
in our model,
the thickness of domain wall $\Delta$ has to be larger than $10^{-3}$ ly.

\item The domain wall happens to travel with the sun. The wall should be big
enough to hold the orbit of the earth around the sun. The distance between the
sun and the earth is $1.5\times10^{8}$ km $\sim10^{-5}$ ly. Thus $\Delta$ has
to be larger than $10^{-5}$ ly.

\item The domain wall happens to travel with the earth. In this case,
$\Delta$ has to be greater than the scale of the earth $\sim10^{4}$km
$\sim10^{-9}$ly.
\end{enumerate}

On the other hand, measurements from SN1987A found $\delta_{\text{SN}}%
=\frac{v_{\nu}-c}{c}<2\times10^{-9}$ \cite{SN1987A}, much smaller than
$\delta_{\text{OPERA}}$ obtained in OPERA. In order to explain the obvious
contradiction, two possible scenarios are discussed below.

\begin{enumerate}
\item The supernova lies in the same $xy$ plane as the earth. In this
scenario, both neutrino and photon travel in $xy$ plane with the same speed.
Careful analysis of the supernova's position and the direction of
the baseline in OPERA has to be carried out to check the possibility of this scenario.

\item The supernova and the earth are not in the same $xy$ plane. By assuming
the supernova is on $z$-axis, the distance from SN1987a to the earth
$L_{\text{SN}}\approx\left\vert z_{f}^{\text{SN}}-z_{i}^{\text{SN}}\right\vert
\sim1.6\times10^{6}$ly. $z_{f}^{\text{SN}}$ is the position of the supernova
and $z_{i}^{\text{SN}}\approx z_{i}^{\text{OPERA}}$ as long as $\Delta$ is
larger than the scale of the earth.\ Thus eqn. (\ref{delta expression})\ yields%
\begin{equation}
\delta_{\text{SN}}\sim\delta_{\text{OPERA}}\frac{\left\vert f\left(
\frac{z_{f}^{\text{SN}}}{\Delta}\right)  -f\left(  \frac{z_{i}^{\text{SN}}%
}{\Delta}\right)  \right\vert }{\frac{L_{SN}}{\Delta}}\sim\delta
_{\text{OPERA}}\frac{\Delta}{L_{SN}},%
\end{equation}
where \ $\delta_{\text{OPERA}}\sim\frac{\sigma^{2}}{M^{4}\Delta^{2}%
}$ and $\left\vert f\left(  \frac{L_{SN}}{\Delta}\right)
\right\vert \le 2 $  have been used.\ The results from OPERA and SN1987a give
$\delta_{\text{SN}}/\delta_{\text{OPERA}}\lesssim10^{-4}$. It implies
$\Delta/L_{\text{SN}}\lesssim10^{-4}$ and $\Delta$\ $\lesssim$\ $10^{2}$ ly.
\end{enumerate}

Finally, for sake of convenience, the neutrino baseline is assumed along
$z$-axis in this letter. However, taking into account the rotation and revolution
of the earth, this is not the case in OPERA experiments. If one is only
interested in order of magnitudes, the above simplification is guaranteed.
Measurements of dependencies of $\delta_{\text{OPERA}}$ on the direction of
the baseline in future experiments could confirm or rule out our proposed model.

In conclusion, we discussed that existence of a domain wall could explain the
recent measurement of superluminal neutrinos in OPERA. We assumed that the
earth lives in a domain wall and the corresponding scalar couples to the
neutrino as well as photon fields through the effective operators in eqn.
(\ref{effective lagrangian}). Once eqn. (\ref{domain condition}) is satisfied,
the observed $\delta_{\text{OPERA}}$ could be explained in the framework.
Furthermore, two scenarios have been proposed to
reconcile OPERA with SN1987a. In one scenario, the supernova and the earth lie
in the same $xy$ plane parallel to the domain wall. Hence neutrinos and photons travel
at the same speed toward the earth after the explosion. In the other scenario,
the thickness of the domain wall is much smaller than the distance between the
supernova and the earth. Then eqn. (\ref{delta expression}) yields
$\delta_{\text{SN}}\ll\delta_{\text{OPERA}}$.

\bigskip
\noindent
{\bf Acknowledgements}
We are grateful to X. Liu for instructive discussions and inspirations.
This work is supported in part by NSFC (Grant No. 11175039 and 11005016) and
Fundamental
Research Funds for the Central Universities (Grant No. ZYGX2009J044 and
ZYGX2009X008).

\end{document}